\documentclass[11pt]{article}
\begin{document}
\thispagestyle{empty}

\begin{center}
{\Large\bf Errors in the Quantum Electrodynamic Mass  
Analysis of Hagelstein and Chaudhary} 
\end{center}
\bigskip

\begin{center}  
{A. Widom\\
{\it Department of Physics, Northeastern University, Boston MA USA}}
\medskip
\end{center}

\begin{center}  
{Y.N. Srivastava\\
{\it Department of Physics \& INFN, University of Perugia PG IT}}
\medskip
\end{center}

\begin{center}  
{L. Larsen\\
{\it Lattice Energy LLC, 175 North Harbor Drive, Chicago IL USA}}
\medskip
\end{center}

\begin{abstract}
Hagelstein and Chaudhary have recently criticized our low energy nuclear reaction 
rates in chemical cells based on our computed electron mass renormalization for 
surface electrons of metal hydride electrodes. They further criticize our electron 
mass renormalization in exploding wire systems which is very strange because mass 
renormalization was {\em never even mentioned} in our exploding wire work. Here we 
show that the calculations of Hagelstein and Chaudhary are erroneous in that they 
are in conflict with the  Gauss law, i.e. they have arbitrarily removed all Coulomb 
interactions in electromagnetic propagators. They have also ignored substantial 
Ampere interactions in favor of computing only totally negligible contributions. 
When the fallacious considerations  of Hagelstein and Chaudhary are clearly exposed, 
it becomes evident that our previous calculations remain valid.  
\end{abstract}

\section{Introduction \label{intro}}
Shortly after claimed evidence for the fusion of deuterium nuclei in room temperature chemical 
cells was withdrawn, the editors of {\em Nature} reported on subsequent 
claims\cite{nature:1989} that large excess heats of reactions were found in chemical cells 
even when the heavy water was replaced with light water. In the last fifteen years or so, many 
experiments have reported low energy nuclear transmutations in chemical cells containing completely 
negligible amounts of deuterium\cite{Miley:1996,Miley:1997,Miley:2005}. To explain both 
the heavy water and light water cases, the present authors investigated alternative 
weak interaction modes of nuclear transmutations as opposed to the claimed strong interaction 
fusion of deuterons. The grounds were in part that strong Coulomb barriers make highly 
improbable the fusion mechanism at room temperature in experimental heavy water systems 
and the lack of deuteron fuel make fusion virtually impossible in light water systems. 
The weak interaction mechanism being investigated involves the reaction 
$$
{\cal X}_{\rm initial}+p^+ +e^-\to n+\nu_e +{\cal X}_{\rm final },
$$
converting protons and electrons into neutrons and neutrinos. The subsequent neutron 
absorption by other nuclei induces nuclear transmutations. The quantum electrodynamic energy 
and mass renormalization required for the initial condensed matter environment 
\begin{math} \cal{X}_{\rm initial} \end{math} to feed electrodynamic energy into the 
neutron producing weak interaction has been previously discussed\cite{Widom:2006}. 

Recently, Hagelstein and Chaudhary have criticized\cite{Hagelstein:2008} our 
calculations of mass renormalization and energy and claimed that our computed mass 
renormalization for surface electrons of metal hydride electrodes is too high. We 
here assert that the calculations of Hagelstein and Chaudhary are erroneous in that 
they are in conflict with the electric Gauss-Coulomb  law. They further criticize our 
mass renormalization in exploding wires which confuses us since {\em we never 
even mentioned mass renormalization} in our exploding wire work\cite{Widom:2007}. 
\medskip \par \noindent 
(i) The error by Hagelstein and Chaudhary in violating the electric Coulomb law 
interaction is discussed in detail in Sec.\ref{eci}. However, let us here note that 
contrary to the Hagelstein-Chaudhary statement that electron energy shifts in the 
MeV energy range are ``unprecedented'', the {\em fact} is that electron energy shifts 
in the MeV energy range are routinely calculated and measured. Consider  the electric 
field (\begin{math} \sim 10^2\end{math} megavolt per centimeter) within the 
surface dipole layer produced when a metal is in contact with an insulator. If the 
insulator were to slide across the metal, then the frictional electron charge transfer 
at the small regions of contact (total area \begin{math} \propto \end{math} to the 
normal force) can support charge separation voltage differences as high as 
\begin{math} \sim \end{math} ten megavolts. The resulting established experimental 
device is a ``Van de Graaf Generator'' and is very well known to produce discharge 
electron energies in the nuclear physics MeV range. While the Van de Graaf mechanism 
is roughly a DC surface charge separation, the surface plasmon excitations discussed 
by Widom and Larsen may be described as an  AC surface charge separation. When 
sufficiently stimulated by a DC current source passing through the metal hydride 
surface of a chemical cell electrode, such AC charge separating plasmon excitations    
may also give rise to MeV electronic energy shifts. Of course, to compute this effect, 
one has to include charge density oscillations in the mean square electric field 
\begin{math} \overline{|{\bf E}|^2} \end{math}. 

{\em Sadly, Hagelstein and Chaudhary throw away those parts of the electric field that 
arise from the charge density}, i.e. those parts of the electric fields which 
can actually describe charge separation. Therein lies the Hagelstein-Chaudhary violation 
of the Coulomb-Gauss law. It would appear that Hegelstein and Chaudhary prefer to 
calculate Coulomb interactions employing the pencil eraser rather than the pencil writing 
point. Arbitrarily erasing the Coulomb barrier between two deuterons has been a common 
technique for some workers proposing the fusion mechanism at room temperature. 
For the weak interaction mechanism, the Coulomb interaction is a help rather than a 
hindrance.  
\medskip \par \noindent 
(ii) The error by Hagelstein and Chaudhary in violating the magnetic Ampere law interaction 
is discussed in detail in Sec.\ref{mai}. The long ranged and unscreened magnetic mean field  
contribution to the single electron annihilation chemical potential is {\em distinct} from 
a field fluctuation mass renormalization.  However, we note that for narrow long current 
channels with currents larger than \begin{math} I_0\approx 17.045089\ {\rm KA}  \end{math}, 
nuclear transmutations have been observed on a variety of systems. These range from narrow wire 
explosions\cite{Wendt:1922,Stephanakis:1972,Young:1977,Bakshaev:2001,Bakshaev:2006,Velikovich:2007,Coverdale:2007}  
to neutron generating lightening bolts\cite{Shah:1985}. The current 
\begin{math} I_0/c=R_{vac}I_0/4\pi =mc^2/e  \end{math} is the measure used by 
Alfv\'en for the electron rest energy.
 
\section{Electric Coulomb Interactions \label{eci}}

The {\em correct} Dirac Hamiltonian for electrons moving in an electromagnetic field when expressed in the 
Coulomb gauge is as follows:
\begin{eqnarray}
{\cal H}=\int \psi^\dagger H\psi d^3{\bf r},
\nonumber \\ 
H=c{\bf \alpha }\cdot {\bf p}+mc^2\beta -e{\bf \alpha }\cdot {\bf A}+e\Phi, \\ 
div{\bf A}=0\ \ \ \ {\rm and}\ \ \ -\nabla^2\Phi =4\pi \rho 
\label{eci1}
\end{eqnarray}
wherein \begin{math} \rho  \end{math} is the charge density and the Coulomb gauge potential 
\begin{math} \Phi  \end{math} obeys the Poisson equation yielding the Coulomb interaction 
\begin{equation}
\Phi ({\bf r},t)=\int \frac{\rho ({\bf r}^\prime ,t)d^3{\bf r}^\prime}{|{\bf r} -{\bf r}^\prime|}.
\label{eci2}
\end{equation}
To arrive at the incorrect Eq.(1) of the Hagelstein-Chaudhary 
manuscript\cite{Hagelstein:2008}, one merely needs to take our correct Eq.(1) and throw away 
(by erasure) the Coulomb interaction Eq.(\ref{eci2}) in the Coulomb gauge. 

Unfortunately, the whole Hagelstein-Chaudhary  manuscript is invalidated by the {\em blunder} of leaving 
out \begin{math} \Phi  \end{math} in their very first equation. In the Hagelstein-Chaudhary 
model world in which the Coulomb gauge photon propagator does not contain 
\begin{math} \Phi \end{math}, one would never find a Coulomb interaction by exchanging a photon 
between charges. This should come as no surprise since in the Hagelstein-Chaudhary model world 
Gauss' law no longer applies and every thing from Hydrogen atoms to condensed matter no longer exist. 
The full photon propagator in the Coulomb gauge contains a term corresponding to 
\begin{math} \Phi  \end{math} fluctuations as is written in any and all of the standard text books on 
quantum electrodynamics\cite{Berestetskii:1997}. In such texts it is clearly explained that both 
longitudinal and transverse fields are required to compute mass renormalizations. 

By keeping the small transverse electric field terms and arbitrarily throwing away the main 
charge density plasmon contribution, Hagelstein and Chaudhary are in gross error in 
evaluating the mass renormalization.

\section{Magnetic Ampere Interactions \label{mai}}

In the Hagelstein-Chaudhary manuscript\cite{Hagelstein:2008}, it was asserted that 
\medskip \par \noindent 
`` ... it was proposed recently by Widom, Srivastava, and Larsen that a very large 
mass shift could be obtained in the strong electromagnetic fields associated with an 
exploding wire experiment. ...''
\medskip \par \noindent 
One might wonder why such an utterly false assertion was made. As any reader who consults 
our paper on exploding wires\cite{{Widom:2007}} can verify, mass renormalization {\em was 
not even mentioned}. Since we have been so blatantly misquoted by Hagelstein and 
Chaudhary, we will briefly review here the essence of what was {\em actually} discussed 
in our exploding wire work. 

For non-relativistic moving charged particles, the effective Lagrangian was deduced by 
Darwin\cite{Darwin:1920}, and is presently discussed in the better electrodynamics 
textbooks\cite{Landau:1975,Jackson:1999}. The Lagrangian has the usual form of a kinetic energy 
minus a potential energy 
\begin{eqnarray}
L=K-U,
\nonumber \\ 
U=\sum_{1\le a < b\le N} \frac{e_ae_b}{r_{ab}},
\nonumber \\ 
K=\sum_{1\le a\le N}\frac{1}{2}m_a|{\bf v}_a|^2+
\sum_{1\le a < b\le N} \frac{e_ae_b}{2c^2r_{ab}}
\left({\bf v}_a\cdot{\bf v}_b+({\bf n}_{ab}\cdot {\bf v}_a)({\bf n}_{ab}\cdot {\bf v}_b)\right),
\label{mai1}
\end{eqnarray} 
wherein \begin{math} {\bf r}_{ab}={\bf r}_a-{\bf r}_b \end{math} and 
\begin{math} {\bf n}_{ab}={\bf r}_{ab}/r_{ab} \end{math}. The effective kinetic energy thus includes 
magnetic Ampere interactions between moving charges.

For the case of currents moving through thin wires, we employed a simple inductive form for the magnetic 
part of the kinetic energy. For a wire of length \begin{math} \Lambda  \end{math}, directed along a unit vector 
\begin{math} {\bf n}  \end{math} and having inductance \begin{math} L  \end{math}, the total kinetic energy is   
\begin{eqnarray}
K_N=\sum_{1\le a\le N}\frac{1}{2}m_a|{\bf v}_a|^2+\frac{1}{2}L\left(\frac{I}{c}\right)^2,
\nonumber \\ 
I=\frac{1}{\Lambda} \sum_{1\le a\le N} e_a{\bf n}\cdot {\bf v}_a, 
\nonumber \\ 
K_N=\sum_{1\le a\le N}\frac{1}{2}m_a|{\bf v}_a|^2+
\frac{L}{2c^2\Lambda^2}\sum_{a,b}e_ae_b{\bf n}\cdot {\bf v}_a{\bf n}\cdot {\bf v}_b.
\label{mai2}
\end{eqnarray} 
The magnetic part of the kinetic energy thereby includes interaction terms between 
the \begin{math} a^{th} \end{math} and \begin{math} b^{th} \end{math} particles. 
The change in kinetic energy which accrues from destroying one charged particle is then 
\begin{eqnarray}
K_N-K_{N-1}=k,
\nonumber \\ 
k=\frac{1}{2}m|{\bf v}|^2+\eta \left(\frac{eI}{c}\right)\frac{{\bf n}\cdot {\bf v}}{c},
\label{mai3}
\end{eqnarray}
wherein the inductance per unit wire length is \begin{math} \eta =(L/\Lambda)\end{math}.
The second term on the right hand side of Eq.(\ref{mai3}) represents the magnetic Ampere  
energy of a single electron moving in a wire with total electronic current \begin{math} I \end{math}. 
A central result of our exploding wire work was written for the magnetic enhancement of the 
kinetic energy as 
\begin{equation}
W_{\rm magnetic}=-\eta mc^2\left(\frac{I}{I_0}\right)\frac{v}{c}.
\label{mai4}
\end{equation}
Many electrons acting cooperatively contribute energy \begin{math} W_{\rm magnetic} \end{math} 
to our weak interaction inverse beta transitions even though only one of those electrons is destroyed.
The energy \begin{math} W_{\rm magnetic} \end{math} is again in the MeV range of nuclear physics 
for the many exploding wire experiments in which neutrons have been observed. 

The above Ampere interaction energies being large, i.e. in the MeV range, were arbitrarily ingnored 
by Hagelstein and Chaudhary who only chose to discuss small and totally negligible magnetic energies. 

\section{Conclusion \label{conc}}

In the Coulomb gauge as with any other gauge, both longitudinal 
and transverse fields enter into the photon propagators and thus into mass renormalization. 
The Hagelstein-Chaudhary rejection of any and all Coulomb interactions is in complete disagreement 
with standard conventional electrodynamics. By contrast,  our mass renormalization (for surface electrons
on metal hydrides) which includes the large Coulomb interaction contribution still stands.

Hagelstein and Chaudhary have also falsely criticized our electron 
mass renormalization in exploding wire systems. 
This criticism is quite bizarre in that we never discussed mass renormalization in our exploding wire work. 
Hagelstein and Chaudhary have also ignored substantial Ampere interactions in favor of computing 
only totally negligible contributions. 

In the light of these erroneous considerations of Hagelstein and Chaudhary, all our previous results 
are still valid.

\end{document}